\documentclass[12pt,aps,nofootinbib]{revtex4}
\usepackage{amsmath}
\usepackage{graphicx}

\def\gsim{\gtrsim}

\newcommand{\Gammamat}{\boldsymbol \Gamma}
\newcommand{\Rmat}{\mathbf R}
\renewcommand{\vec}[1]{\boldsymbol #1}

\DeclareMathOperator{\Trace}{Tr}

\begin{document}

\title{Functional RG for imbalanced  many-fermion systems}
\author{Boris Krippa$^{1}$}
\affiliation{$^1$School  of Science and Technology,
Nottingham Trent University, NG1 4BU, UK}
\date{\today}

\maketitle

Fermionic pairing occurs in many physical systems from molecular physics to quark matter at finite density. The underlying dynamical mechanisms share a common feature of Cooper instability leading to a rearrangement of the ground state and associated spontaneous symmetry breaking.

Here we focus on the asymmetric ultracold atomic Fermi gas of two fermion flavours, which realizes a highly tunable system of strongly interacting fermions which is tuned due to  Feshbach resonance, which allows to control the interaction strength between two different species of fermions and explore the BEC-BCS crossover in a wide range of  parameters. Another tunable parameter (in asymmetric systems) is the population imbalance which can be used to probe how stable the superfluid phase is. The problem was studied   in \cite{CC}  where it has been shown that in the BCS limit the system with the chemical potential mismatch $\delta\mu$ undergoes first order phase transition to a normal phase at $\delta\mu_{c} = 0.71 \Delta_0$ where $\Delta_0$ is the gap at zero temperature for balanced system. Since then the imbalanced many-fermion systems have been studied in a number of papers. However, most theoretical studies have been performed in the framework of the mean-field (MF) type of approaches which  may not be sufficiently reliable in  providing quantitative answers. In many cases the effects of quantum fluctuations turn out to be important.

The convenient tool to study pairing phenomena in  many-fermion systems is provided by the  Functional Renormalisation Group \cite {Wet}  (FRG) where the effects of quantum fluctuations are included in a consistent and reliable way. The  FRG approach makes use of  the Legendre transformed  effective action: $\Gamma[\phi_c]=W[J]-J\cdot \phi_c$, where $W$ is the usual partition function in the presence of an external source $J$. The action functional $\Gamma$ generates the 1PI Green's functions and it reduces to the effective potential for homogeneous systems. In the FRG  one introduces an artificial renormalisation group flow, generated by a momentum scale $k$ and we define the
effective action by integrating over components of the 
fields with $q \gsim k$. The RG trajectory then interpolates between the 
classical action of the underlying field theory (at large $k$), and the 
full effective action (at $k=0$). 

The evolution equation for $\Gamma$ in the ERG has a 
one-loop structure  and can be written as
\begin{equation}
\partial_k\Gamma=-\frac{i}{2}\,\Trace \left[
(\Gammamat ^{(2)}_{BB}-\Rmat_B)^{-1}\,\partial_k\Rmat_B\right]
+\frac{i}{2}\,\Trace \left[
(\Gammamat ^{(2)}_{FF}-\Rmat_F)^{-1}\,\partial_k\Rmat_F\right].
\label{eq:Gamevol}
\end{equation}
Here $\Gammamat ^{(2)}_{FF(BB)}$ is the matrix containing second
functional derivatives of the effective action with respect to the
fermion (boson) fields and $\Rmat_{B(F)}$ is a matrix containing the
corresponding boson (fermion) regulators which must vanish when the running scale approaches zero.
  A $2\times 2$ matrix structure 
arises for the bosons because we treat $\phi$ and $\phi^\dagger$ as
independent fields in order to include the number-violating condensate.
A similar structure also appears for the fermions. By inserting the
ansatz for $\Gamma$ into this equation one can turn it into a set of
coupled equations for the various couplings.

We study a system of fermions with population imbalancies interacting through an attractive 
two-body point-like potential and consider pairing between the fermions with different flavours assuming that the interaction between the identical ones is negligible. We take as our starting point  the $s$-wave scattering of two distinguishable fermions in vacuum with a $T$-matrix 
determined by the scattering length $a$.

Since we are interested in the appearance of a gap in the fermion
spectrum, we need to parametrise our effective action in a way that
can describe the qualitative change in the physics when this occurs.
A natural way to do this is to introduce a boson field whose vacuum 
expectation value (VEV) describes the gap and so acts as the 
corresponding order parameter.  At  
the start of the RG flow, the boson 
field is not dynamical and is introduced through a 
Hubbard-Stratonovich transformation of the four-fermion pointlike interaction.
As we integrate out more and more of the fermion degrees of freedom by 
running $k$ to lower values, we generate dynamical terms in the bosonic
effective action.

The ansatz for $\Gamma$ used here is a generalisation of the ansatz used in \cite{Kri1} for a balanced many-fermion system
\begin{eqnarray}
\Gamma[\psi,\psi^\dagger,\phi,\phi^\dagger,\mu,k]&=&\int d^4x\,
\left[\phi^\dagger(x)\left(Z_\phi\, i \partial_t 
+\frac{Z_m}{2m}\,\nabla^2\right)\phi(x)-U(\phi,\phi^\dagger)\nonumber\right.\\
&&\qquad\qquad+ \sum_{i=1}^{i=2}\psi^\dagger\left( Z_\psi (i \partial_t+\mu_i)
+\frac{Z_{M_i}}{2M_i}\,\nabla^2\right)\psi\nonumber\\
&&\qquad\qquad\left.- g\left(\frac{i}{2}\,\psi^{\rm T}\psi\phi^\dagger
-\frac{i}{2}\,\psi^\dagger\psi^{\dagger{\rm T}}\phi\right)\right],
\label{eq:Gansatz}
\end{eqnarray}
Here $M_i$ and $m$ are masses of fermions and composite boson. All renormalisation factors, couplings and chemical potentials run with the scale $k$. The term containing the boson chemical potential is quadratic in $\phi$ so it can be absorbed into effective potential $U$ and the Yukawa coupling is assumed to describe the decay (creation) of a pair of nonidentical fermions. Due to $U(1)$ symmetry the effective potential depends only  on the combination $\phi^{\dagger}\phi$. We expand the potential $U(\rho)$ near its minima  and keep terms up to order $\rho^3$.
\begin{equation}
U(\phi,\phi^\dagger)= u_0+ u_1(\rho-\rho_0)
+\frac{1}{2}\, u_2(\rho-\rho_0)^2 + \frac{1}{6}\, u_3(\rho-\rho_0)^3 + . . . ,
\label{eq:potexp}
\end{equation}
where $\rho = \phi^{\dagger}\phi$. We  assume $Z_{\psi_i} = Z_{M_i} = 1$ and neglect running of Yukawa coupling.
One notes that the expansion near minimum of the effective potential (either trivial or nontrivial), being  reliable in the case of  second order phase transition, may not be
 sufficient for the first order one. It is worth emphasizing that  the CC limit related transition from the superfluid phase to a normal one is of the first order so 
that a reliability of the expansion needs to be verified.  However, as we will discuss below, at small/moderate asymmetries even a simple ansatz for the effective action  the effective potential expanded up to the third order in the field bilinears gives a reasonable description of the corresponding phase diagram and provides a clear evidences that the phase transition is indeed of first order.
 
At the starting scale the system is in a symmetric regime with a trivial minimum so that $u_{1}(k)$ is positive. At some lower scale $k = k_{c}$ the coupling $u_{1}(k)$ becomes zero and the system undergoes a transition to the broken phase with a nontrivial minimum and develops the energy gap.

 In our RG evolution we have chosen the trajectory when chemical potentials run in the broken phase and the corresponding particle densities $n_i$ remain fixed so that we define
"running Fermi-momenta" for two fermionic species as $p_i = \sqrt{2 M_i \mu_i}$. It is convenient to work with the total chemical potential and their difference so we define
\begin{equation}
\mu = \frac{\mu_1 + \mu_2}{2}; \qquad \delta = \frac{\mu_1 - \mu_2}{2}
\end{equation}
and assume that $\mu_1$ is always larger then $\mu_2$. Calculating corresponding functional derivatives, taking the trace and performing a contour integration results in the following flow equation for the effective potential
\begin{eqnarray}
\partial_k U
&=&-\,\frac{1}{2 Z_\psi}\int\frac{d^3{\vec q}}{(2\pi)^3}\,\frac{E_{1F} + E_{2F}}
{\sqrt{(E_{1F} + E_{2F})^2+ 4 g^2\rho}}\,(\partial_k
R_{1F} + \partial_k R_{2F})\nonumber\\
\noalign{\vskip 5pt}
&&+\,\frac{1}{2Z_\phi}\int\frac{d^3{\vec q}}{(2\pi)^3}\,
\frac{E_{BR}}{\sqrt{E_{BR}^2-V_B^2}}
\,\partial_kR_B,\label{eq:potevol}
\end{eqnarray}
where
\begin{equation}
E_{BR}(q)=\frac{Z_m}{2m}\,q^2 + U'' \rho + U' + R_B(q,k),\qquad
V_B = U''\rho,
\end{equation}
and 
\begin{equation}
E_{iF} \equiv E_{iF}(q,k,p_i)=\epsilon_{i}(q)-\mu_i+R_{iF}(q,p_i,k),\qquad \epsilon_{i}(q) = q^2/2 M_i.
\end{equation}
Here we denote $U' = \frac{\partial U}{\partial \rho}$ and $U'' = \frac{\partial^2 U}{\partial \rho^2}$ etc.

One notes that the position of the pole in the fermion loop integral which defines the corresponding dispersion relation is given by

\begin{equation}
q_0 = \frac{E_{2F} - E_{1F} \pm \sqrt{(E_{2F} + E_{1F})^2+ 4 \Delta^2}}{2},
\end{equation}
where $\Delta^2 =  g^2 \rho$ is the square of the pairing gap.

In the physical limit of vanishing scale this dispersion relation indicates a possibility of the gapless exitation in asymmetric many-fermion systems (much discussed Sarma phase \cite{Sar}). The gappless
 exitation occurs  at $\frac{\Delta}{\delta} <$ 1.  As we will show below,  this condition is never fulfilled so that Sarma phase does not occur.
We note, however, that this conclusion is valid at zero temperature case and can be altered at finite temperature where the possibility for the Sarma phase  still exists.
The corresponding bosonic exitations are just gapless "Goldstone" bosons as it should be.

In order to follow the evolution at constant density and running chemical potential we  define the total derivative
\begin{equation}
d_k=\partial_k+(d_k\mu)\,\frac{\partial}{\partial\mu} + (d_k\rho)\,\frac{\partial}{\partial\rho}  , 
\end{equation}
where $d_k\mu=d\mu/dk$,   $d_k\rho=d\rho/dk$. Applying this to effective potential,  
demanding that $n$ is constant ($d_k n=0$) gives
the set of the flow equations 
\begin{eqnarray}
2Z_{\phi}\,d_k\rho - \chi d_k \mu
&=&\left.\frac{\partial}{\partial \mu}
\Bigl(\partial_k \bar U\Bigr)\right|_{\rho=\rho_0},\\
\noalign{\vskip 5pt}
d_k u_0+n\,d_k\mu
&=&\left.\partial_k U\right|_{\rho=\rho_0},\\
\noalign{\vskip 5pt}
-u_2\,d_k\rho + 2Z_\phi\,d_k\mu
&=&\left.\frac{\partial}{\partial \rho}
\Bigl(\partial_k U\Bigr)\right|_{\rho=\rho_0},\\
\noalign{\vskip 5pt}
 d_k u_2 -u_3 d_k \rho - d_k\mu\beta&=&\left.\frac{\partial^2}{\partial \rho^2}
\Bigl(\partial_k U\Bigr)\right|_{\rho=\rho_0},\\
\noalign{\vskip 5pt}
\frac{1}{2}\chi' d_k \mu + d_k Z_\phi + \frac{1}{2}\beta d_k \rho &=&-\,\frac{1}{2}\left.\frac{\partial^2}{\partial \mu\partial\rho}
\Bigl(\partial_k U\Bigr)\right|_{\rho=\rho_0},\\
\noalign{\vskip 5pt}
-\beta' d_k \mu + d_k u_3 - u_4 d_k \rho &=&\left.\frac{\partial^3}{\partial\rho^3}
\Bigl(\partial_k U\Bigr)\right|_{\rho=\rho_0}
\end{eqnarray}
where we have defined
\begin{equation}
\chi = \frac{\partial^{2} U}{\partial \mu^2},  \qquad  \chi' = \frac{\partial^{3} U}{\partial \mu^2 \partial \rho}, 
\qquad  \beta = \frac{\partial^{3} U}{\partial \mu \partial \rho^2}, \qquad  \beta' = \frac{\partial^{4} U}{\partial \mu \partial \rho^3}
\end{equation}

The left-hand sides of these equations contain a number of higher order terms  such as $u_4$, $\chi$, $\chi'$, $\beta$, $\beta'$.  The scale dependence of   these couplings is obtained from evolution with fermion loops only.

The driving terms in these evolution equations are given by appropriate 
derivatives of Eq.~(\ref{eq:potevol}). In the symmetric phase we evaluate 
these expressions at $\rho=0$. The driving term for the chemical potential evolution 
vanishes in this case, and hence $\mu$ remains constant. In the broken phase 
we keep $\rho$ non-zero and set $u_1=0$. 

One notes that neglecting the effect of bosonic fluctuations leads to the mean-field expression for the effective potential \cite{Kri2}

Our approach can be applied to any type of the imbalanced many-fermion system but for a concretness we use a parameter set relevant to nuclear matter: $M_1 = M_2 = 4.76$ fm$^{-1}$, $p_{1(2)} \simeq 1$ fm$^{-1}$ and large fermion-fermion scattering length ($a \>> 1$) fm.
 The initial conditions for $u'$s and $Z$ can be obtained by differentiating the expression for the effective potential at the starting scale $k = k_{st}$ and setting the parameter $\rho$ to zero.

Now we turn to the results.  First we note that the system undergoes the transition to the broken phase at critical scale $k_{c} \simeq \frac{p_1 + p_2}{2}$. Its value slowly decreases when the asymmetry is increased while keeping the total chemical potential fixed. We found that the value of the $k_{kr}$ is practically insensitive to the starting scale provided the scale is chosen to be larger than 10 fm$^{-1}$.  The position of the minimum of the effective potential in the unitary regime changes rather slowly with increasing $\delta\mu$  until the chemical potential mismatch  reaches some critical value $\delta\mu_{c} = 0.67\mu$. At $\delta\mu$ larger then   $\delta\mu_{c}$ the minimum of the effective potential drops to zero  thus indicating first order phase transition similar to the CC limit, obtained in the  weak coupling regime.

\begin{figure}
\begin{centering}
\includegraphics[width=10cm]{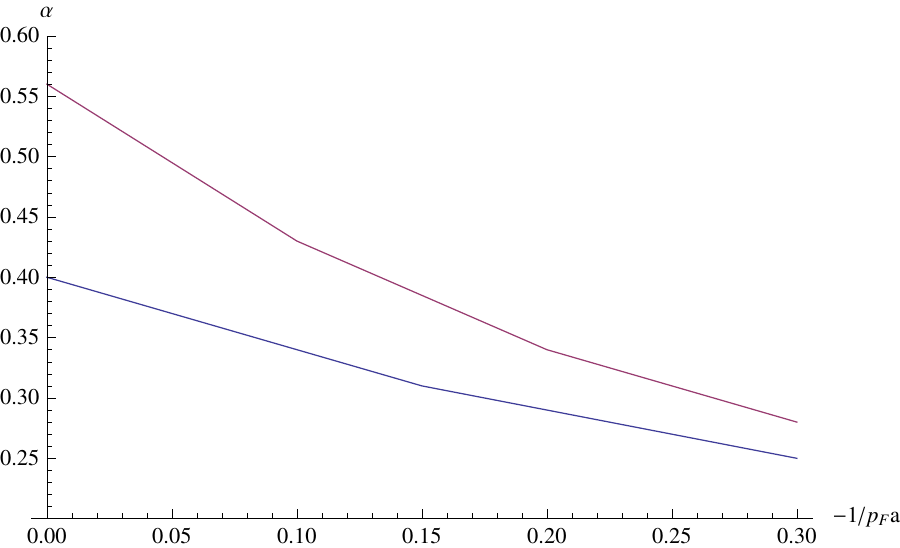}
\par\end{centering}

\caption{Phase diagram as a function of -1/$p_F a$ and polarisation with $p_F$ corresponding to the fermions with a larger density. The upper curve (red online) is the result 
of the calculations and lower curve (blue online) corresponds to the fit of experimental data from \cite{Hul}}

\end{figure}

It is worth emphasizing that the effective potential, expanded to third order in $\rho$ exhibits two minima as it should be for the first order phase transition. We have picked up the minimum at 
$\rho = \rho_0$ as it corresponds to lower energy.
In Fig.1 we show the results for the critical line, separating the gapped and normal phases as a function of the dimensionless parameter $1/p_F a$, where $p_F$ corresponds to the
state with larger density and particle asymmetry density  $\alpha = \frac{n_1 - n_2}{n_1 + n_2}$.
 The experimental data are from  \cite{Hul}. The lower curve  is the  fit of the data from \cite{Hul}. 
Our theoretical curve approaches  the  fit  with decreasing $p_F |a|$ although always lies  above the experimental data thus indicating the room for a further improvement of the ansatz.  One notes, that at any value of -1/$p_F a$ the phase transition always takes place when $\frac{\Delta}{\delta}$ is greater then one so that the  condition required for the Sarma phase is never reached. One can therefore conclude that, at least at $T = 0$, Sarma phase never occurs so that the phase  transition is indeed of first order otherwise we would find  that at some point the ratio $\Delta/\delta$ becomes less then one. Physically it means that the system must be viewed as an inhomogeneous mixture of the gapped and normal phases, as suggested in \cite{Bed}.\\
The higher order  couplings bring in the corrections on the level of 18-20 $\%$ so the expansion of the effective potential near minimum converges reasonably well. Certainly,  in order to improve the description of the experimental data,  the full solution for the unexpanded potential is required but a qualitative conclusion about phase transition being of first order will remain the same regardless of the way the effective potential is treated.
It is worth emphasizing that the unified description of all possible phases (superfluid, mixed and normal) as the function of the imbalance would require a substantial modification of the formalism. The reason is that the particle-hole channel leading to some corrections in the superfluid phase ( Gorkov-Melik-Barkhudarov (GMB) corrections \cite{Gor}) becomes a "main player" in the mixed and normal phases so that the reliable  description of  all possible phase boundaries requires an inclusion of the particle-hole channel in a consistent and  nonperturbative way. This is technically very challenging problem which has never been solved for the case of imbalanced many-fermion systems.

We show on Table 1 the results of the calculations for the superfluid gap in the limit of small density imbalance $\alpha = 0.03$ in comparison with the experimental data from \cite{Shin1}.
 As in the case of the phase diagram the theoretical points are not  far from the experimental data but still lie above them indicating that higher order terms should be included in
 our truncation
 for the effective action to achieve better agreement with the data.\\
\begin{table}[ht]
 \caption{Superfluid gap}
 \centering 
\begin{tabular}{c c c c} \hline\hline
 $1/p_{F}a$ & $\Delta$(exp) & $\Delta$(calc)\\ [0.5ex]	
 \hline 0&0.44&0.55\\
 -0.25&0.22&0.27 \\
  [1ex]
 \hline
 \end{tabular} 
\label{table:nonlin}
 \end{table}

 We have also calculated the critical value of the chemical potential mismatch $\delta\mu_c$ with parameters typical for neutron matter (scattering length $a_{nn} \simeq -18.6$) fm. Again at large
 enough $\delta\mu > \delta\mu_{crit}$ the pairing is disrupted and the system undergoes to a normal phase. The value of  $\delta\mu_c$ can be important for the phenomenology of neutron stars
 because the
 transport properties of the  normal and superconducting phase are very different \cite{Gez1}. Our calculations gives the value  $\delta\mu_c = 0.33
\mu$ to be compared with  the QMC based
 results $\delta\mu_c = 0.27\mu_c$ \cite{Gez2}.

One notes that the FRG approach has also been applied  to  imbalanced many-fermion system  in the recent paper \cite {Paw1} where the unexpanded effective potential  has been used. Similar to our paper, the conclusion about a nonexistence of the Sarma phase at zero temperature has been made. The quantitative results for $\delta\mu_{c}$ obtained in our paper differ from those from \cite {Paw1} by approximately 15$\%$  which can be related to the higher order corrections.  In general, one can conclude that, in spite of  a relative simplicity of the assumed ansatz for the effective action  FRG provides a good starting point for a  reasonable description of the phase diagram of asymmetric many-fermion systems. The phase transition is found to be of first order in agreement with the other theoretical results \cite{Pit} and the Sarma phase never occur for this system (at zero temperature) which means that the system should be interpreted as an inhomogeneous mixture of the gapped and normal phases.

An important step forward would be an  inclusion of the 
fermion-fermion interaction in the particle-hole (ph) channel leading to the GMB corrections \cite{Gor} which are crucial for a realistic description of the mixed and normal phases. The FRG based studies of the GMB corrections have been performed in \cite{Die} for the case of the balanced many-fermion systems. A generalisation of the approach developed in \cite{Die} to the imbalanced systems is highly nontrivial and requires a serious technical and computational efforts.

Now let us make some remarks concerning the studies of the cold atom dynamics in a more general context of strongly interacting many-fermion systems, like nuclei. There are indeed many analogies between cold atoms and nuclear matter like an existence of Cooper pairs and Fermi-surfaces. However, there are important differences which must be kept in mind. First of all, nuclear interactions are not pointlike and, moreover, exhibit a repulsion on the short distances. It leads to rather different behavior of the pairing gap when the fermionic density is increased such that the "atomic gap" grows with the density and "nuclear gap" saturates. It means that some exact results obtained in cold atom system like for example the short range behavior of the two-fermion wave function described by the Tan contact \cite{Tan} cannot a'priori be expected to hold for nuclear systems. Second, unlike cold atoms nucleons  are effective degrees of freedom so that the description of nuclei in terms of nucleons and mesons makes sense up to a certain momentum scale. It implies that, starting from  some "critical" density (chemical potential) the description of the system in terms of quarks and gluons looks more adequate, bringing in typical QCD issues like confinement which is very hard to model using cold atoms.

\section{acknowledgement}
The author is grateful to M. Birse and N. Walet for valuable discussions.

\end{document}